\documentclass[2p,review]{elsarticle}
\usepackage{geometry}
\usepackage{lineno,hyperref}
\journal{Journal of Fluids and Structures}
\usepackage{amsmath,todonotes}
\usepackage[nameinlink,noabbrev]{cleveref}
\newcommand{\kindex}[2]{\ensuremath{{#1}_{\scalebox{0.5}{#2}}}}
\newcommand{\U}{\textrm{U}}
\newcommand{\Uinf}{\kindex{\U}{$\infty$}}
\usepackage[locale=UK,
number-mode=math,
unit-mode=text,
per-mode=symbol,
number-unit-product=\ ,
retain-zero-exponent=true]{siunitx}
\DeclareSIUnit{\pixel}{px}
\DeclareSIUnit{\fps}{fps}

\usepackage{etoolbox} 
\robustify{\kindex}
\usepackage{paralist}

\graphicspath{{../figurematter/latex/figs/}{../figurematter/plots/}}

\renewcommand{\hspace}[1]{}
\newcommand\Rey{\mbox{\textit{Re}}~}

\usepackage[normalem]{ulem}

\begin{document}

\begin{frontmatter}

\title{The dynamics and timescales of static stall}

\author[mainaddress]{S\'ebastien Le Fouest}
\author[mainaddress]{Julien Deparday\fnref{newaff}}
\author[mainaddress]{Karen Mulleners\corref{mycorrespondingauthor}}
\cortext[mycorrespondingauthor]{Corresponding author}
\ead{karen.mulleners@epfl.ch}

\address[mainaddress]{Institute of Mechanical Engineering, \'Ecole Polytechnique F\'ed\'erale de Lausanne (EPFL), CH-1015 Lausanne, Switzerland}
\fntext[newaff]{Present affiliation: Institute for Energy Technology, Eastern Switzerland University of Applied Sciences (OST), CH-8645 Rapperswil, Switzerland}

\begin{abstract}
Airfoil stall plays a central role in the design of safe and efficient lifting surfaces.
We typically distinguish between static and dynamic stall based on the unsteady rate of change of an airfoil's angle of attack.
Despite the somewhat misleading denotation, the force and flow development of an airfoil undergoing static stall are highly unsteady and the boundary with dynamic stall is not clearly defined.
We experimentally investigate the forces acting on a two-dimensional airfoil that is subjected to two manoeuvres leading to static stall: a slow continuous increase in angle of attack with a reduced pitch rate of \num{1.3e-4} and a step-wise increase in angle of attack from \SIrange{14.2}{14.8}{\degree} within \num{0.04} convective times.
We systematically quantify the stall reaction delay, or the timespan between the moment the blade exceeds its critical static stall angle and the onset of stall, for many repetitions of these two manoeuvres.
The onset of flow stall is marked by the distinct drop in the lift coefficient.
The reaction delay for the slow continuous ramp-up manoeuvre is not influenced by the blade kinematics and its occurrence histogram is normally distributed around \num{32} convective times.
The static reaction delay is compared with dynamic stall delays for dynamic ramp-up motions with reduced pitch rates ranging from \numrange{9e-4}{0.14} and for dynamic sinusoidal pitching motions of different airfoils at higher Reynolds numbers up to \num{1e6}.
The stall delays for all conditions follows the same power law decrease from \num{32} convective times for the most steady case down to an asymptotic value of \num{3} convective times for reduced pitch rates above \num{0.04}.
Static stall is not phenomenologically different than dynamic stall and is merely a typical case of stall for low pitch rates where the onset of flow separation is not promoted by the blade kinematics.
Based on our results, we suggest that conventional measurements of the static stall angle and the static load curves should be conducted using a continuous and uniform ramp-up motion at a reduced frequency around \num{1e-4}.
\end{abstract}


\end{frontmatter}

\section{Introduction}

Flow separation and stall play a central role in the design of lifting surfaces for a wide range of applications such as rotary and fixed wing aircraft, wind turbines, etc. \cite{McCullough_Gault_1951, McCroskey_Fisher_1972, Leishman_2002}.
Stall is a commonly encountered, mostly undesired, condition that occurs when the angle of attack of an airfoil exceeds a critical angle.
We typically distinguish between static and dynamic stall based on the rate of change of the airfoil's angle of attack \cite{Ericsson.1988}.
The distinction is rather qualitative, as there is no universal criterion to assess whether a motion can be considered either static or dynamic.
The denotation of static stall is highly misleading for two reasons:
\begin{inparaenum}[(i)]
\item an airfoil can not stall unless it moves past its critical stall angle and
\item the flow and force development during the transition from an attached to a separated flow state are inherently unsteady.
\end{inparaenum}
The temporal evolution of aerodynamic loads acting on an airfoil undergoing stall at extremely low pitch rate is often overlooked, as most attention is devoted to dynamic motions.

Literature regarding dynamic stall was initially motivated by helicopter rotor aerodynamics and flutter \cite{Mccroskey1981,Carr1977,Smith.2020e08}, and received renewed interest due to problems related to gust interactions \cite{Farnsworth.2020,Jones.2020}.
The main parameter governing flow unsteadiness related to the kinematics of a pitching airfoil is the reduced pitch rate $k$ defined as:
\begin{equation}
k = \frac{c \dot{\alpha}}{2\Uinf}\quad,
\end{equation}
where $c$ is the airfoil chord, $\dot{\alpha}$ is the pitch rate in radians per second, and $\Uinf$ is the free stream velocity.
This parameter represents the ratio of the kinematic to the convective timescales of the flow.
The reduced pitch rate can be thought of as a phase lag between the blade kinematics and the surrounding fluid's response, resulting from the inertial effects \cite{Leishman2000}.
For high enough pitch rates, the blade experiences a significant lift overshoot compared to the static case, and stall onset is delayed to an angle of attack beyond the critical stall angle.
The additional lift is attributed to the formation, growth and shedding of large-scale dynamic stall vortices \cite{Carr1977,McAlister1978}.
The angular delay of flow separation is considered one of the classical hallmarks of dynamic stall \cite{Mccroskey1981}.
From a timing perspective however, high pitch rates promote flow separation and reduce the blade's reaction time relative to the static case \cite{Mulleners2012, Kissing2020}.

The reaction time is a measure of the time the blade takes to stall after its angle of attack exceeds the critical stall angle.
This timespan follows a power law decay for increasing reduced pitch rates, reaching a plateau for reduced pitch rates above \num{0.04} \cite{Deparday2019,Eldredge:2018cu}.
The minimum value for reaction time is attributed to the vortex formation time.
The dynamic stall vortex requires a certain convective time to form before massive flow separation can occur, typically between \num{3} and \num{5} convective times for pitching airfoils \cite{Gharali2013,Dabiri2009}.
For extremely low pitch rate values, blade kinematics cease to promote flow separation and the reaction time is expected to reach a maximum value which is not yet well defined.
The key differences in the temporal evolution of static and dynamic aerodynamic loads remain to be formulated, and arguably static stall should be regarded as a general case of stall for extremely low values of the pitch rate.

Conventionally, we consider stall to be static when the airfoil's kinematics are slow enough to avoid delaying full flow detachment past the airfoil's critical stall angle.
This angle is measured by making sub-degree angle of attack increments, allowing the flow around the airfoil to fully develop before further motion is imposed.
The last angular position before a loss in lift is observed is considered the critical stall angle.
This value plays a crucial role in characterising the airfoil's performance for dynamic motions and is a key parameter in semi-empirical models for dynamic stall \cite{Sheng2008,Leishman1989}.
Guidelines that characterise the relationship between reduced pitch rate and the temporal occurrence of stall are relevant to accurately determine the critical stall angle of a given airfoil.

We experimentally investigate the transient development of aerodynamic forces during what is typically considered as conventional static stall.
We approach the static stall limit in two different ways: by increasing the angle of attack in small discrete steps and by slowly but continuously increasing the angle of attack.
The systematic acceleration and deceleration related to the stepwise increase of the angle of attack is more likely to disturb the flow than a continuous, extremely slow ramp-up.
The time resolved lift response to the two types of quasi-steady motions is compared.
Specific focus is directed towards the identification of successive stages in the flow development and the statistical analysis of the timescales associated with the different flow development stages.
We quantify the reaction delay between the time when the blade exceeds its static stall angle and the occurrence of stall, determine the limiting values for extremely low and high pitch rates, and compare the results with the stall delays measured for various dynamic motions.
The main objective is to characterise the influence of the reduced pitch rate on the characteristic timescales of an airfoil undergoing stall, and identify qualitative properties that help qualify a motion to be static or dynamic.

\section{Experimental setup}																					%
\begin{figure}
\centering
\includegraphics{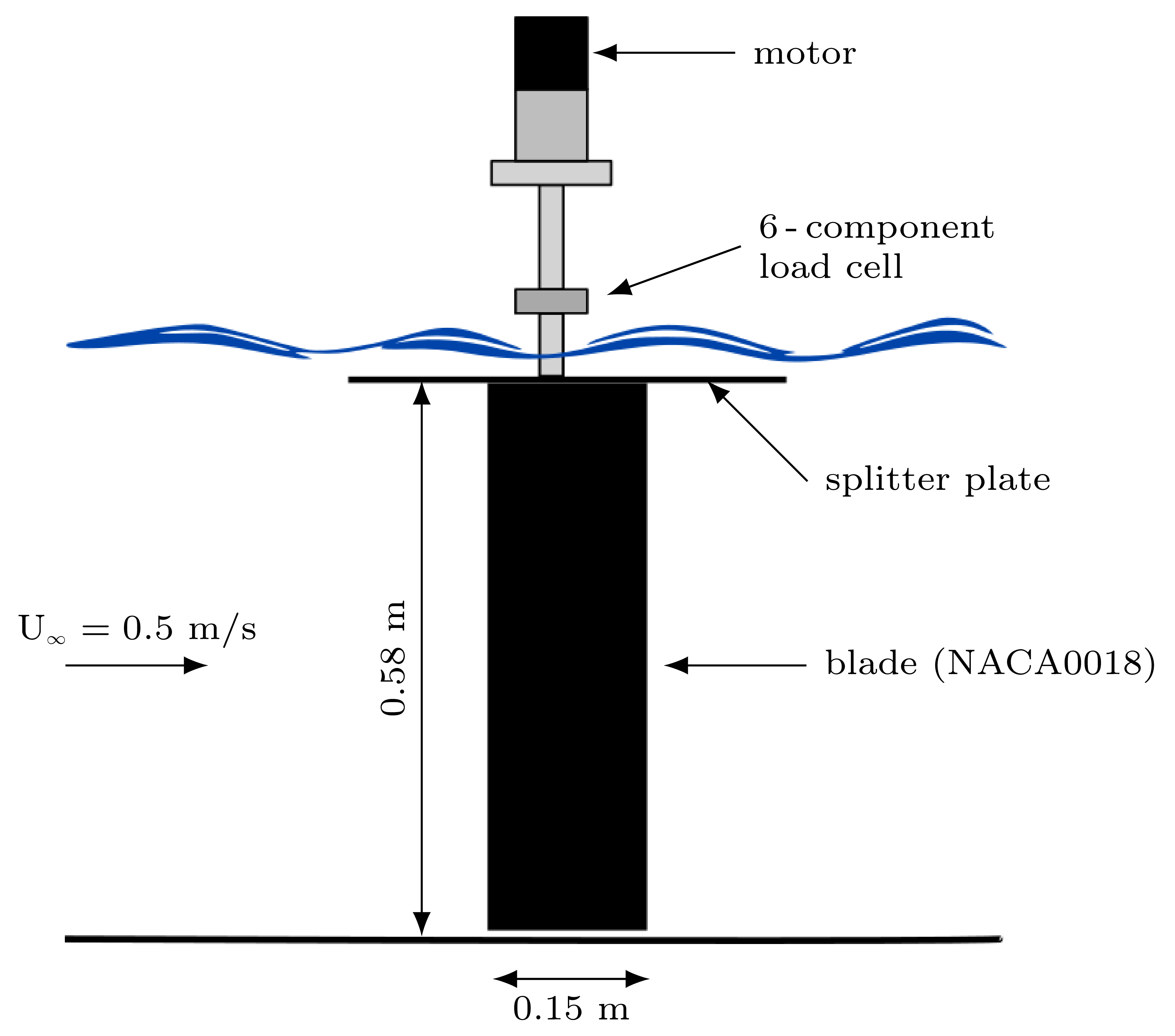}
\caption{Schematic representation of the two-dimensional NACA0018 blade mounted vertically in a free surface recirculating water channel with a cross section of \SI{0.6x0.6}{\meter}.}
\label{fig:expsetup}
\end{figure}

A NACA0018 profile is vertically suspended in a recirculating water channel with fully-transparent acrylic windows and a test section with dimensions \SI{0.6 x 0.6 x 3}{\metre}, illustrated in \cref{fig:expsetup}.
The airfoil's chord is c = \SI{0.15}{\metre} and the span is close to the height of the test section: S = \SI{0.58}{\metre}.
The bottom of the profile is flush with the bottom transparent wall, and a splitter plate is mounted on the top to reduce free surface effects on the airfoil.
The blade was 3D printed as a single piece using PLA, sanded with ultra-fine \num{1000}-grit paper, and covered with a thin layer of epoxy resin to rigidify the blade and ensure it is watertight.
The rotation happens about the quarter-chord axis, and it is driven by a \SI{10}{\milli\meter} square-sectioned stainless steel shaft that is tightly inserted through the blade's span.
The incoming flow is set at \Uinf = \SI{0.50}{\metre\per\second}, resulting in a Reynolds number of Re = \num{7.5e5}.
Forces are measured with a six degrees of freedom load cell (ATI Nano 25) placed at the interface between the shaft of the airfoil and the motor shaft.
Force data were recorded with a sampling frequency of \SI{1000}{\hertz}, a sensing range of \SI{125}{\newton} and a resolution of \SI{0.02}{\newton}.
Output from the load cell was transmitted to a computer using a data acquisition system (National Instruments).
Buoyancy forces and load cell factory offset were measured in the water channel without flow for the whole range of angles of attack investigated.
These forces were subtracted from the force data to obtain the aerodynamic loads.
The force data was filtered using a second-order low-pass filter with the cut-off frequency of \SI{30}{\hertz}.
This frequency is multiple orders of magnitude larger than the pitching frequency and \num{48} times larger than the expected post-stall vortex shedding frequency based on a chord-based Strouhal number of \num{0.2} \cite{Leishman1989}.

Two quasi-steady manoeuvres that lead to the occurrence of what can be considered conventional static stall are investigated: a slow continuous ramp-up motion and a step-wise increase.
For the first type of manoeuvre, we continuously increased the angle of attack from $\alpha=\ang{8}$ up to $\alpha=\ang{18}$ with a constant pitch rate of \SI{0.05}{\degree\per\second}, which corresponds to a reduced pitch rate of $\num{1.3e-4}$.
For the second type of manoeuvre, we performed a step-wise increase in the angle of attack from \SIrange{14.2}{14.8}{\degree} within \num{0.04} convective times at \SI{50}{\degree\per\second}, which corresponds to a reduced pitch rate of $k= \num{1.3}$.
The airfoil was rotated around its quarter-chord axis using a stepper motor with a $\ang{1.8}$ step angle and $1:25$ planetary gearbox reduction, resulting in a position accuracy of $\ang{0.072}$.
The initial blade position was calibrated using a small angle of attack sweep between \ang{-5} and \ang{5} to find the angle that resulted in zero net lift.
The pitch rate for the continuous ramp-up motion was selected by progressively reducing the pitch rate until we systematically measured no significant lift overshoot at the onset of stall.
Each motion was repeated for approximately \num{100} runs to allow for a statistical analysis of the transient aerodynamic load fluctuations occurring during stall development.
For the step-wise, full flow reattachment was insured between runs by returning the blade to $\alpha=\ang{8}$ and pitching up to its starting position at $\alpha=\ang{14.2}$ with the same slow pitch rate used for the continuous ramp-up motion.
The stall angle was determined based on the data collected with the ramp-up motion and found to be $\kindex{\alpha}{ss}=\ang{14.2}$.
This information was used to select the start angle for the step-wise motion.
Ramp-up motions with higher pitch rates ranging from \SIrange{0.3}{53.3}{\degree\per\second}, corresponding to reduced pitch rates ranging from \numrange{1e-3}{1.4e-1}, were performed to obtain an overview of the influence of the pitch rate on the delay of stall with respect to the static case.
For all experiments, we recorded loads for \SI{5}{\second} or \SI{16.7}{} convective times prior to the start of the manoeuvre, continuously during the manoeuvre, and for another \SI{5}{\second} after the manoeuvre was completed.
The load cell recorded all three components of the force and the moments around all three spatial axes, but we will focus our analysis and discussion on the lift measurements.
In most practical applications, the loss of lift due to a transition from attached to fully separated flow raises the most immediate concern.
The characteristic timescales that we extract based on the lift response are the result of changes in the flow development which would equally affect other forces and moments, such as the drag or the pitching moment.

\section{Results}																					%

\begin{figure}[tb]
\centering
\includegraphics{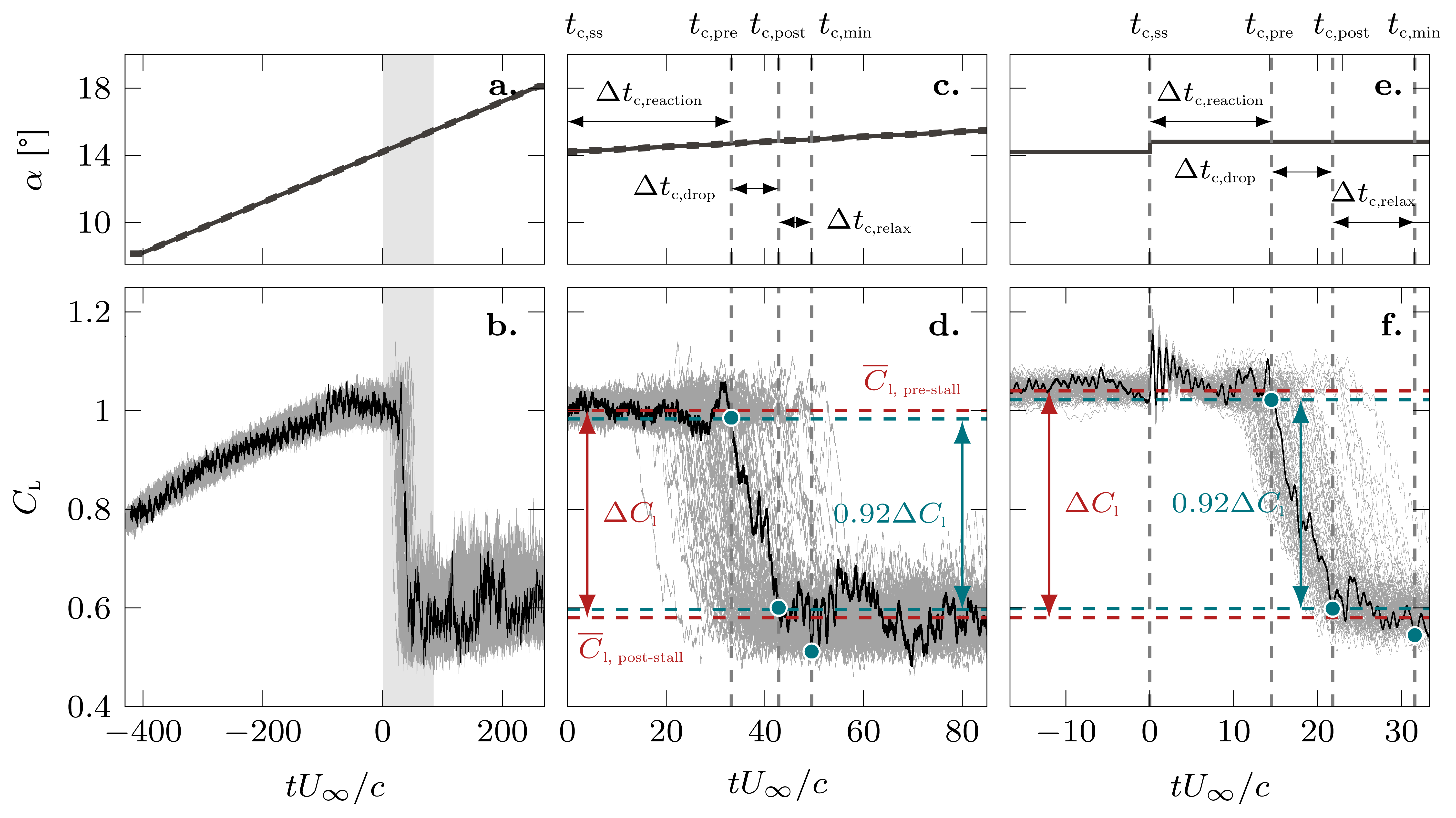}
\caption{
(a) Variation of the angle of attack during the continuous ramp-up manoeuvre.
(b) The corresponding lift responses for 94 repetitions.
(c,d) Close up views of the transient region corresponding to the shaded areas in the left plots.
(e) Variation of the angle of attack during the step manoeuvre.
(f) The lift response to the step-wise increase in the angle of attack for 100 repetitions.
Convective time $t\Uinf/c = 0$ corresponds to the time when the airfoil's angle of attack exceeds the critical angle of \ang{14.2}.
The thicker black lines in the bottow row show the lift coefficient for a single run to highlight the characteristics of the individual responses.
Pre-stall and post-stall reference levels are indicated by the red dashed lines.
The green dashed lines indicate the upper and lower threshold values used to identify the timing of the stall stage.
The start of the lift drop is \kindex{t}{c,pre} is defined as the time at which the lift has dropped more than \SI{4}{\percent} of the total lift drop $\Delta\kindex{C}{l}$.
The end of the lift drop is \kindex{t}{c,post} is defined as the time at which the lift has dropped more than \SI{96}{\percent} of the total lift drop.
}
\label{fig:allforces}
\end{figure}

The temporal evolution of the lift coefficient in response to the slow continuous ramp-up and the step manoeuvre are shown in \cref{fig:allforces}.
The timing is indicated in terms of convective time defined as $\kindex{t}{c}=t\Uinf/c$, relative to $\kindex{t}{c}=0$ when the blade's angle of attack exceeds \ang{14.2}.
The value of \ang{14.2} is considered the critical angle of attack above which the blade will always stall even if no further motion is imposed.
Across the entire range of experiments conducted with this blade under the given flow conditions, \ang{14.2} was the lower bound at which stall would consistently occur.
Results are insensitive to the exact choice of critical angle, as moving it up or down one sub-degree would result in a homogeneous shift in timing for all experimental runs.
A close-up view of the lift drop during stall for the continuous ramp-up in \cref{fig:allforces}d facilitates the comparison between the continuous ramp-up and the step manoeuvre.
Each line in \cref{fig:allforces}b,d,f represents a different run.
A single run is highlighted in black for both experiments to show the characteristics of the response.

In the continuous ramp-up experiment (\cref{fig:allforces}a-d), the lift does not immediately drop once the blade exceeds the critical angle and remains at an approximately constant value for a certain time before dropping.
The time delay between the moment at which the critical angle is exceeded and the moment at which the lift starts to drop differs for each run.
We call this time delay the reaction delay and refer to the period covering the reaction delay as the holding stage.
Although the blade's angle of attack can increase by as much as \ang{0.6} during the holding stage for some runs, the blade does not produce additional circulation and the lift coefficient fluctuates steadily around \num{1} during this stage.
For the highlighted run, the holding stage lasts about \num{33} convective times, followed by the drop stage where the lift coefficient falls from $\kindex{C}{L} = $\numrange{1.05}{0.6} in $9$ convective times.
A third characteristic time instant is identified as the first local lift minimum below the average post-stall level.
At this point, the flow is considered to be in a fully developed stalled state.
The fully developed post-stall stage is dominated by periodic load fluctuations due to characteristic bluff-body-like vortex-shedding \cite{Huang1995,Huang2000}.

For the step-wise increase in angle of attack (\cref{fig:allforces}e,f), the blade is static for the first \SI{5}{\second} or \num{16.7} convective times of the load recording during which the lift coefficient fluctuates around \num{1.05}.
The blade is then subjected to a step-wise increase of \ang{0.6} beyond the critical angle of attack.
The origin on the convective timescale marks again the time at which the blade exceeds the critical stall angle.
The loads for all runs show a highly repeatable inertial response to the step manoeuvre characterised by lift fluctuations at \SI{4}{\hertz} over \num{5.5} convective times.
The fluctuations progressively decay and the lift returns to the same level as before the step manoeuvre.
Similarly to the response to the continuous ramp-up motion, the lift does not collapse immediately after the angle is increased beyond the critical angle but remains at its pre-stall value for a time duration that varies for all runs.
For the highlighted run, the lift coefficient has a last local peak above the pre-stall level around \num{14} convective times after the blade manoeuvre is conducted, followed by a drop from \numrange{1.05}{0.6} in \num{9} convective times.
About \num{32} convective time after the step manoeuvre, the flow reaches a post-stall stage dominated by large periodic load fluctuations.

For both manoeuvres, we distinguish three flow stages that characterise the transient flow development from fully attached to a fully stalled condition: a holding stage, a drop stage, and a relaxation stage.
The time delays associated with these stages are indicated in \cref{fig:allforces} and are defined as follows:
\begin{enumerate}
\item the reaction delay defined as $\kindex{\Delta t}{c,reaction} = \kindex{t}{c,pre} -  \kindex{t}{c,ss}$, where $\kindex{t}{c,pre}$ is the time when the lift drops below an upper threshold limit, and $\kindex{t}{c,ss}$ is when the blade exceeds its critical stall angle of \ang{14.2},
\item the drop time delay $\kindex{\Delta t}{c,drop} = \kindex{t}{c,post} - \kindex{t}{c,pre}$ where $\kindex{t}{c,post}$ is the time when the lift coefficient drops below a lower threshold limit,
\item the relaxation delay $\kindex{\Delta t}{c,relax} = \kindex{t}{c,min} -  \kindex{t}{c,post}$, where $\kindex{t}{c,min}$ is the time when the lift reaches its first local post-stall minimum.
\end{enumerate}
The upper and lower threshold levels are reference values that are determined with respect to the average pre-stall and post-stall limits and are fractions of total lift drop.
Using thresholds at each boundary of the drop section, instead of using local extrema values, reduces the sensitivity of timescales to the large fluctuations occurring in the reaction and relaxation stages.
Here, the upper threshold level was set \SI{4}{\percent} of the total lift drop $\Delta\kindex{C}{l}$ below the average pre-stall value \kindex{\overline{C}}{l,pre-stall}.
The lower threshold level was set \SI{96}{\percent} of the total lift drop below the average pre-stall value \kindex{\overline{C}}{l,pre-stall} or \SI{4}{\percent} of the total lift drop above the average post-stall value \kindex{\overline{C}}{l,post-stall} .
The percentages of the lift drop for the threshold limits were selected following an iterative procedure that aimed at maximising the drop stage length while avoiding outliers issuing from fluctuations at the edges of the drop stage.
The drop stage analysis was completed following the IEEE standards of a negative-going transition \cite{IEEEstd}.
A general idea of the influence of the choice of the threshold values on the results is provided by displaying results obtained with lower and higher threshold limits in the following figures.

\begin{figure}
\centering
\includegraphics{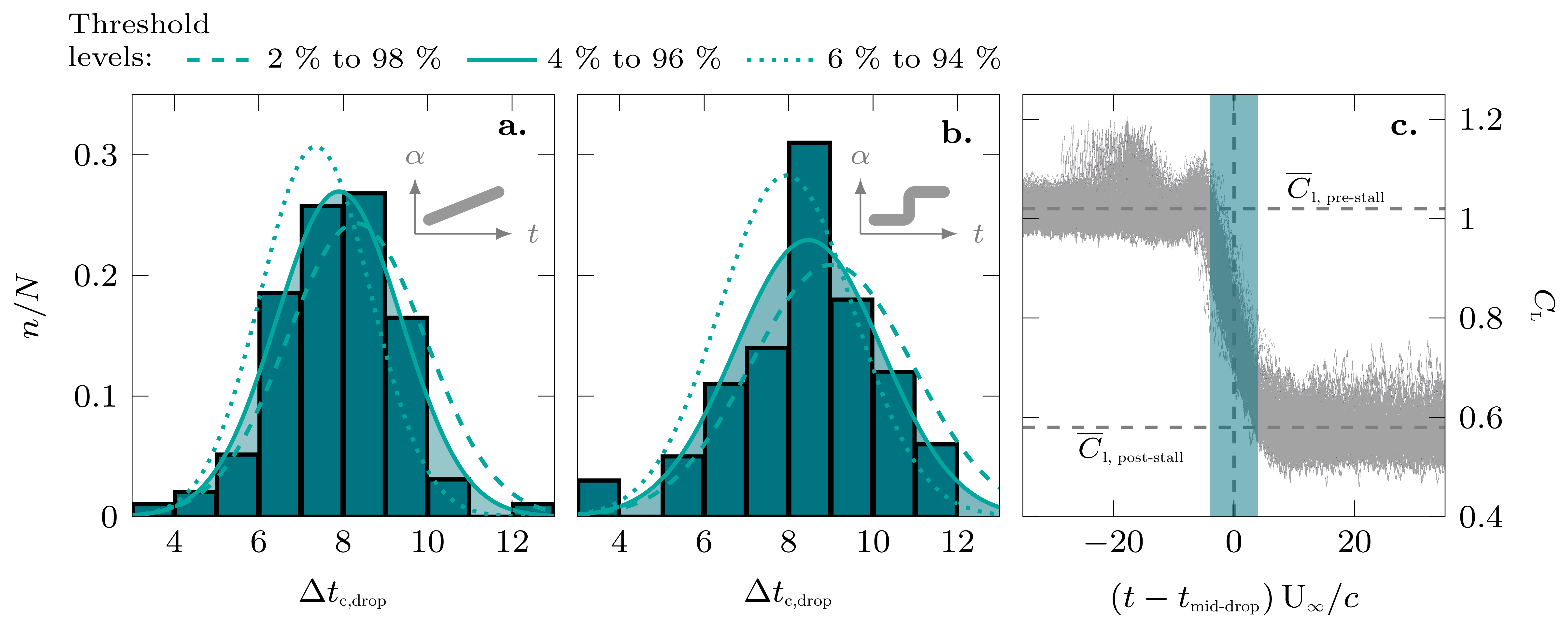}
\caption{Drop time $\kindex{\Delta t}{c,drop}$ occurrence histogram for all runs of (a) the slow continuous ramp-up and (b) the step manoeuvre based on threshold levels of \SI{4}{\percent} and \SI{96}{\percent} to identify the start and end of the lift drop.
The solid line represents the fitted normal distribution corresponding to the histogram.
The dashed and dotted lines indicate how the normal distributions shift when varying the threshold values.
(c) Temporal evolution of the the lift coefficient for all the runs of both the slow ramp-up and step-wise experiment shifted along the time axis with respect to the middle of the drop delay.
The shaded area indicates the overall average drop delay of \num{8} convective times.
}
\label{fig:drophists}
\end{figure}

The distributions of the extracted drop times $\kindex{\Delta t}{c,drop}$ for both manoeuvres are compared in \cref{fig:drophists}a,b.
Both cases display a normal distribution centred around \num{8} convective times.
The lift drop rate is constant for both experiments and yields a highly repeatable portion of the stall transient.
The selected upper limit end lower limit threshold values represent a trade off between spanning the largest possible region for the drop stage, while limiting the spread of the reaction time occurrence histograms.
Narrowing the threshold percentages (\SIrange{6}{94}{\percent}) reduces the spread but will artificial increase the repeatability of the reaction and relaxation stages by accounting for a greater portion of the repeatable drop stage in the neighbouring stages.
Widening the threshold percentages (\SIrange{2}{98}{\percent}) increases the sensitivity of the drop stage duration to fluctuations occurring during the end of the reaction and the beginning of the relaxation stage, resulting in an increased standard deviation of the distribution.
The numerical values depend slightly on the selected thresholds for the identification of the lift drop start and end, but the distributions of the drop times consistently show a normal distribution and the lift evolution during the drop stage is highly repeatable across all repetitions.
The self-similarity of the lift response during the drop time is clearly visualised in \cref{fig:drophists} where all the lift responses for both cases are shown on top of each other shifted in time with respect to the middle of the drop time.

\begin{figure}
\centering
\includegraphics{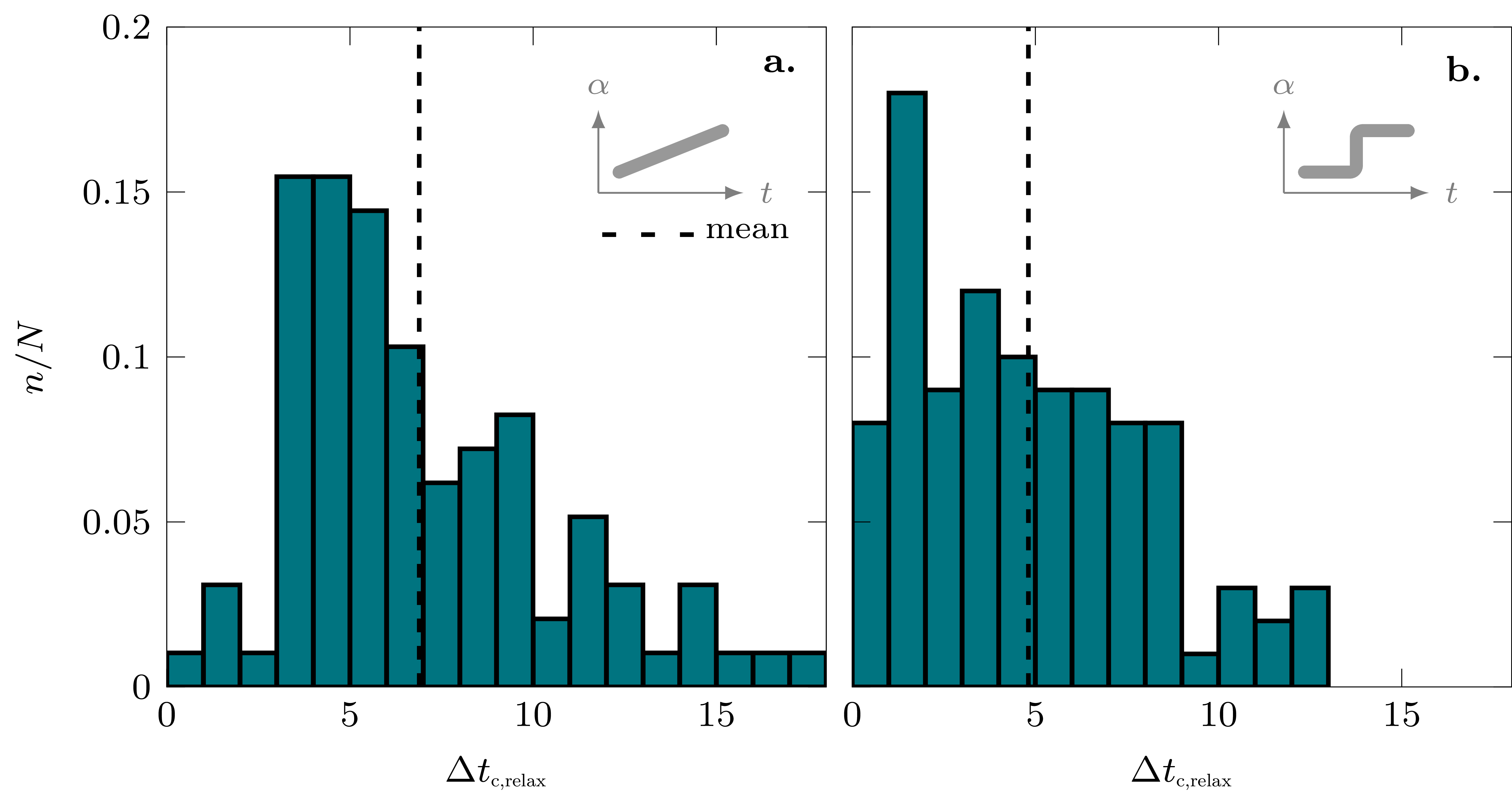}
\caption{Relaxation delay $\kindex{\Delta t}{c,relax}$ occurrence histogram for all runs of (a) the slow continuous ramp-up and (b) the step manoeuvre.
}
\label{fig:tminhists}
\end{figure}

The distributions of the relaxation times $\kindex{\Delta t}{c,relax}$ are compared for both manoeuvres in \cref{fig:tminhists}.
The relaxation time $\kindex{\Delta t}{c,relax}$ shows similar characteristics for both motions investigated: a skewed distribution with a mean value around \num{5}.
The lift coefficient has a standard deviation close to \SI{10}{\percent} of its mean value in the post-stall stage.
These significant fluctuations complicate the identification of the actual relaxation time, as several local minima could be considered as the first post-stall minimum.
Further analysis of the influence of fluctuations on flow detachment will clarify the role that instabilities play in post-stall relaxation.

\begin{figure}
\centering
\includegraphics{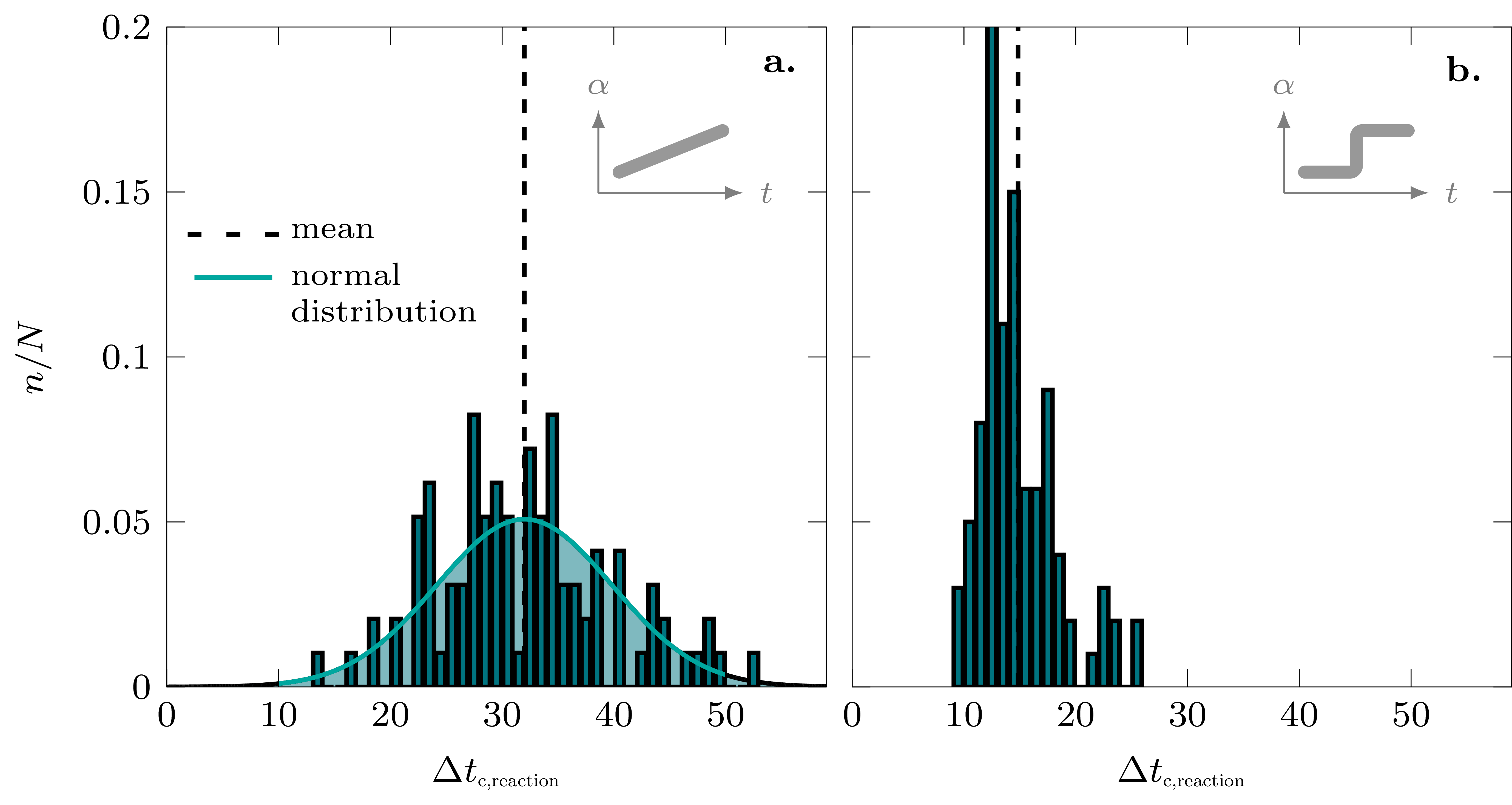}
\caption{Reaction delay \kindex{\Delta t}{c,reaction} occurrence histogram for all runs of (a) the slow continuous ramp-up and (b) the step manoeuvre.
}
\label{fig:tmaxhists}
\end{figure}

The distributions of the reaction times \kindex{\Delta t}{c,reaction} are compared for both manoeuvres in \cref{fig:tmaxhists}.
The reaction time follows a standard normal distribution with an average of \num{32} convective times and a standard deviation of \num{8} convective times for the slow continuous ramp-up manoeuvre (\cref{fig:tmaxhists}a).
Periodic peaks spaced by approximately \numrange{4}{5} convective times suggest that pre-stall fluctuations cause periodically returning conditions that are favourable for the boundary layer to start separating.
The reaction time for the step manoeuvre (\cref{fig:tmaxhists}b) follows a skewed normal distribution with a mean value of \num{14} convective times and a standard deviation of \num{4} convective times.
The skewness, shorter time delay, and narrow spread relative to the continuous ramp up motion are attributed to the step angle of attack increase to beyond the static stall angle.
The sudden and fast motion disturbs the surrounding flow, yielding increased load fluctuations compared to the slow continuous ramp-up manoeuvre.
This unsteadiness promotes full flow detachment and creates a bias in the reaction time occurrence, increasing the repeatability between runs.

\begin{figure}[tb]
\centering
\includegraphics{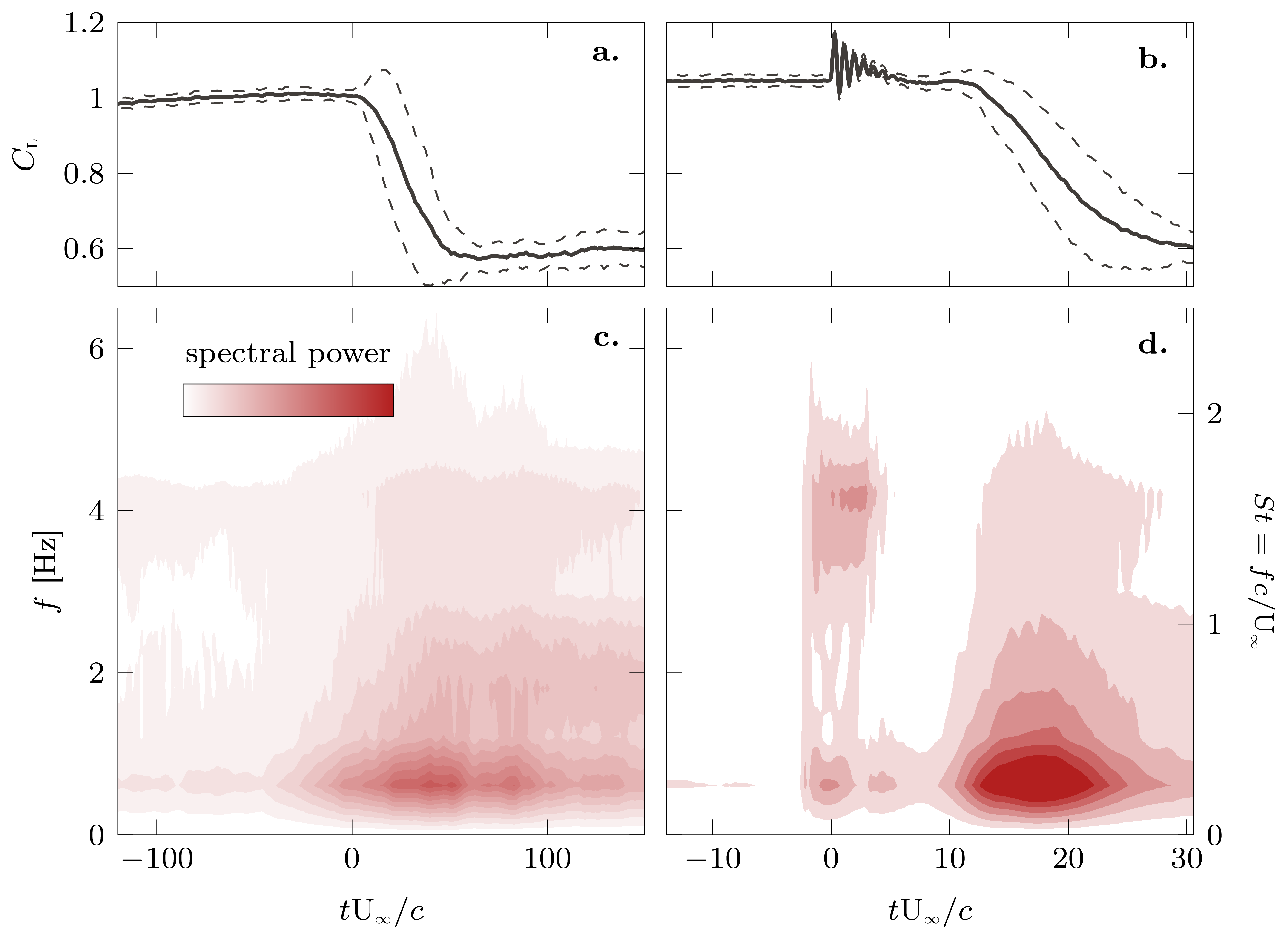}
\caption{Ensemble-averaged temporal evolution of lift (top) and of the amplitude spectrum of lift (bottom) for (a,c) the transient portion of the slow continuous ramp-up manoeuvre and (b,d) the step manoeuvre.
}
\label{fig:freqplot}
\end{figure}

To estimate the dominant frequencies of instabilities and assess their role in the occurrence of stall, we calculated a time-frequency plot for both manoeuvres.
The dominating frequencies of the load fluctuations were quantified by completing a fast Fourier transform (FFT) on the load coefficient.
The FFT was calculated using a sliding time window over the whole time domain for each individual repetition of both manoeuvres.
The window width was set to \num{5.5} convective times, which corresponds to the time duration of the load transient that followed the step manoeuvre.
The ensemble-averaged temporal evolution of the amplitude spectra are presented in \cref{fig:freqplot} for both manoeuvres.
The ensemble-averaged temporal evolution of lift was included to facilitate the comparison between time-scales, load fluctuations, and dominant frequencies.
The timing is indicated in terms of convective time relative to $\kindex{t}{c}=0$ when the blade's angle of attack exceeds \ang{14.2}.
The blade undergoes load fluctuations with a frequency of \SI{4.2}{\hertz} immediately after the step around $\kindex{t}{c}=0$, as highlighted in \cref{fig:freqplot}b.
This response to a single-point excitation on the blade apparatus is equivalent to a modal test.
The highest energy peak in the amplitude spectrum, around \SI{4.2}{\hertz}, corresponds to the natural frequency of the system.
Further peaks in the vicinity of \SI{4.2}{\hertz} are assumed to correspond to structural vibrations.

High energy peaks are observed around \SI{0.65}{\hertz} in the transient lift drop region at the time when lift peaks for both manoeuvres: near \num{32} convective times for the slow continuous ramp-up and near \num{18} convective times for the step manoeuvre.
The expected vortex shedding frequency is calculated based on a Strouhal number $St=\num{0.20}$ found experimentally for a NACA0018 operating at a Reynolds number of $Re = \num{7.5e5}$ \cite{Yarusevych2011}.
The chord was chosen as characteristic length scale for vortex shedding.
The projected chord length is more significant to characterise the interplay between vortices in a fully developed wake.
The formation of stall vortices was found to occur much closer to the blade in the early stages of wake development.
Additionally, our experiment is at a relatively low Reynolds number, so viscosity plays a more important role, reducing vortex formation length to near the blade \cite{Ansell2020,Mulleners2017a}.
Following this argumentation, the expected vortex shedding frequency $\kindex{f}{s}$ is:
\begin{equation}
	\kindex{f}{s}=\frac{\Uinf St}{c} = \SI{0.67}{\hertz}\quad .
\end{equation}

This frequency corresponds to the dominating load fluctuation frequencies observed around \SI{0.65}{\hertz} in the stall transient and post-stall stages for both manoeuvres \cref{fig:freqplot}.
A load fluctuation frequency of \SI{0.65}{\hertz} corresponds to temporally spaced lift peaks of about \num{5} convective times.
This value also corresponds to the spread in the drop time histogram (\cref{fig:drophists}).
The temporal spacing between the peaks of the reaction time histogram for the slow continuous ramp-up manoeuvre (\cref{fig:tmaxhists}a) was also around around \num{5} convective times.
The onset of stall occurs at periodically returning conditions after a randomly distributed number of cycles when the flow is no longer influenced by the blade kinematics.
The amplitude peaks around \SI{4.2}{\hertz} in the stall transient and post-stall stages highlight structural vibrations.
This analysis highlights the fact that flow unsteadiness plays a central role in the timing of the transient lift drop at stall.
The frequency of load fluctuations explains the periodicity of the reaction time distribution and confirms the high repeatability of the drop time.

\begin{figure}
\centering
\includegraphics{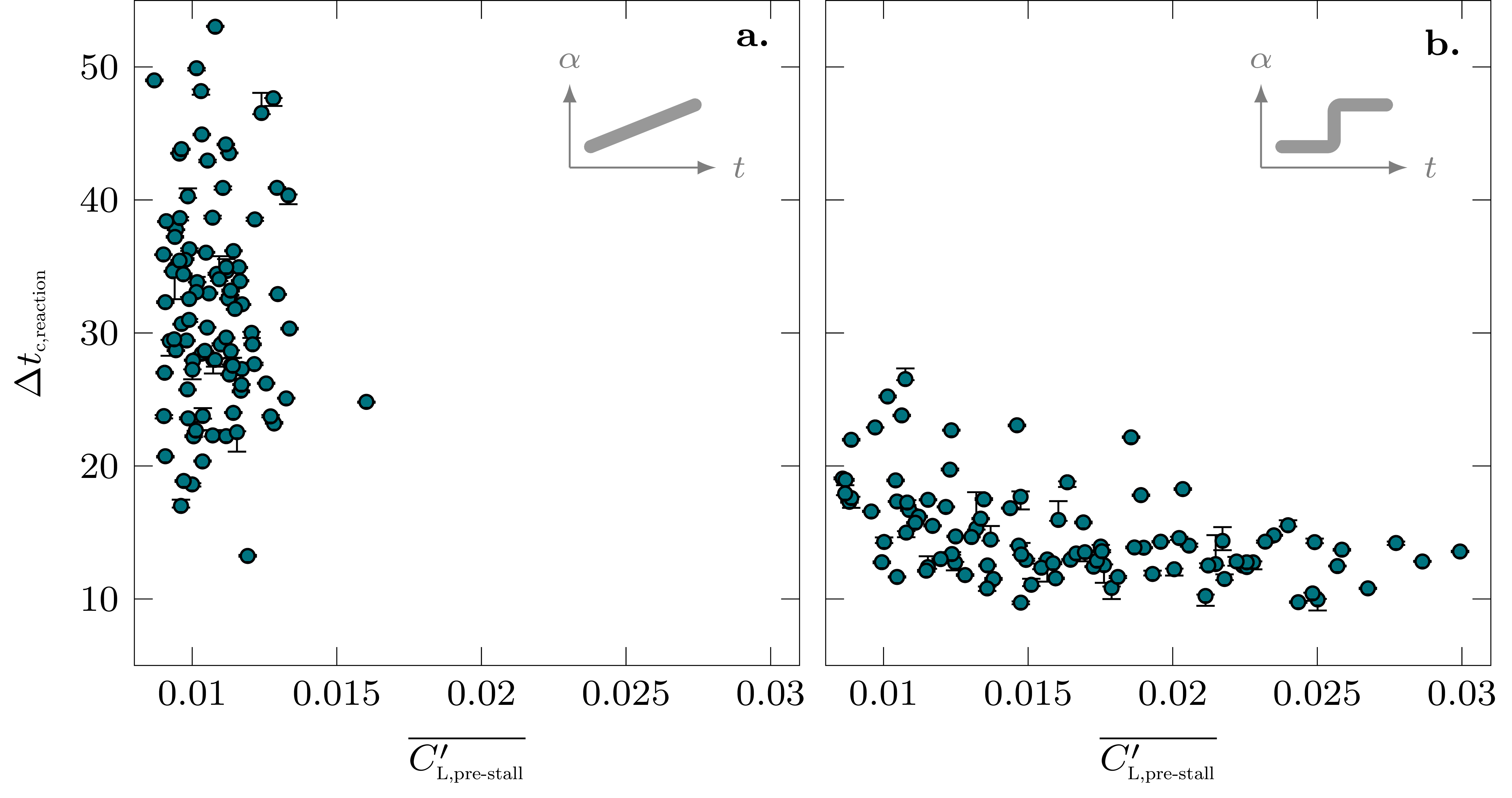}
\caption{Reaction time \kindex{\Delta t}{c,reaction} against load fluctuations for all runs of (a) the slow continuous ramp-up manoeuvre and (b) the step manoeuvre.
Error bars represent the difference between results obtained with threshold levels \SIrange{6}{94}{\percent} and \SIrange{2}{98}{\percent} to identify the start and end of the drop stage.
}
\label{fig:tmaxflucts}
\end{figure}

The reaction and relaxation delays varied significantly between individual runs for both manoeuvres.
To assess the correlation between load fluctuation and the reaction and relaxation times, the magnitude of the load fluctuations was quantified.
To obtain the magnitude of load fluctuations, the mean lift coefficient was computed on a sliding window with a \SI{5} convective time width over the full time domain.
The standard deviation was systematically calculated relative to the local mean $\overline{\kindex{C}{L}}$.
The lift fluctuation magnitude in the holding stage $\overline{\kindex{C'}{L,pre-stall}}$ is computed as the average of the local standard deviations:
\begin{equation}
\overline{\kindex{C'}{L,pre-stall}} = \frac{1}{\kindex{N}{pre-stall}} \sum_{n=1}^N \left(\kindex{C}{L,n} - \overline{\kindex{C}{L,n}}\right)^2
\end{equation}
where $N$ is the number of points in the holding stage, $\kindex{C}{L,n}$ is the local lift coefficient and $\overline{\kindex{C}{L,n}}$ is the mean lift coefficient across the local window.
The lift fluctuation magnitude during the drop stage $\overline{\kindex{C'}{L,drop}}$ was calculated with an analogous expression, replacing $\kindex{N}{pre-stall}$ by $\kindex{N}{drop}$.

\begin{figure}
\centering
\includegraphics{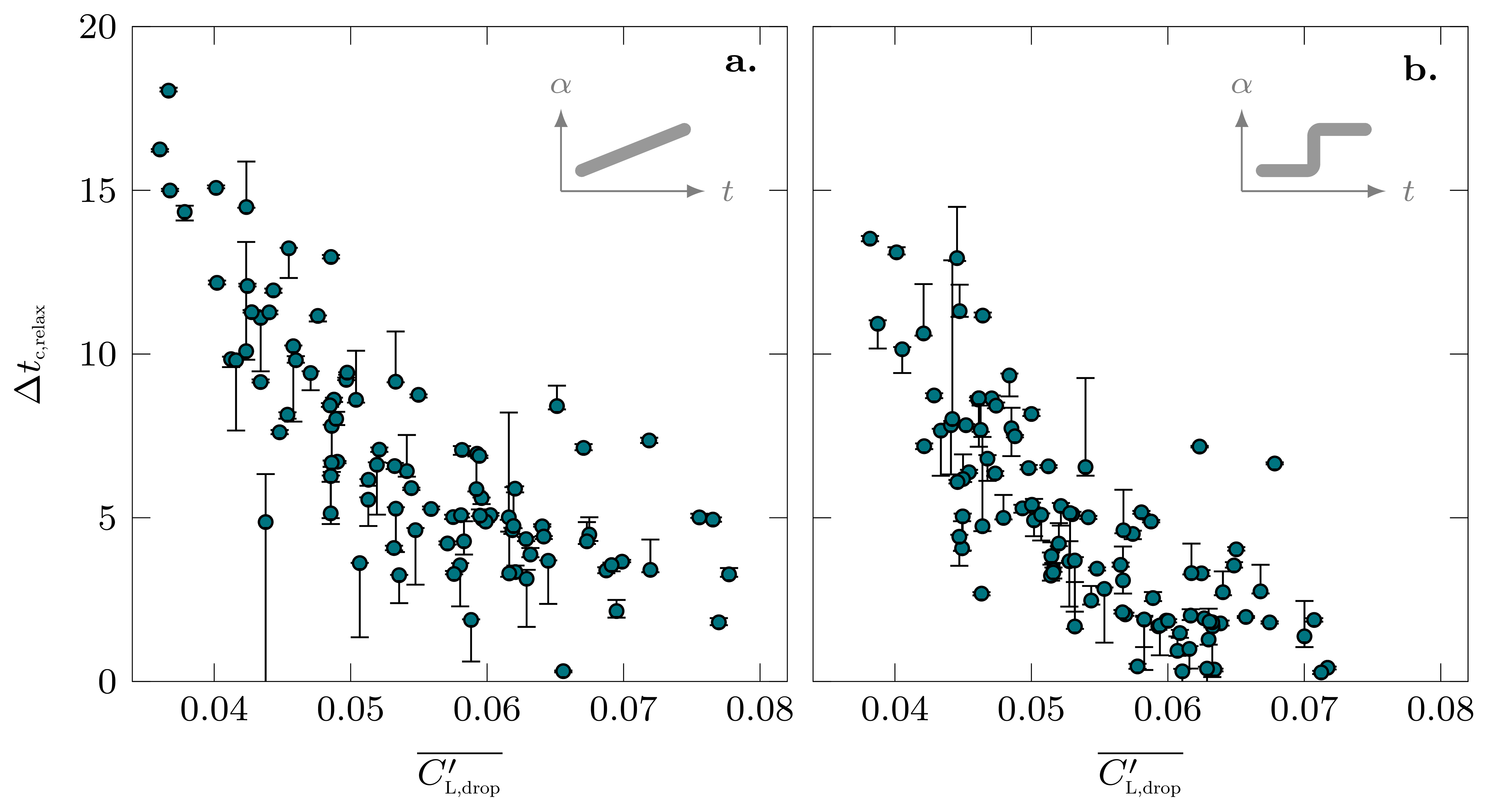}
\caption{Relaxation time $\kindex{\Delta t}{c,relax}$ against load fluctuations for all runs of (a) the slow continuous ramp-up manoeuvre and (b) the step manoeuvre.
Error bars represent the difference between results obtained with threshold levels \SIrange{6}{94}{\percent} and \SIrange{2}{98}{\percent} to identify the start and end of the drop stage.
}
\label{fig:tminflucts}
\end{figure}

The reaction delay \kindex{\Delta t}{c,reaction} is presented against load fluctuations in the holding stage for both manoeuvres in \cref{fig:tmaxflucts}.
Load fluctuations for the slow continuous ramp-up manoeuvre are confined between \num{0.09} and \num{0.013}, which is around \SI{1}{\percent} of the local mean lift coefficient.
There is no apparent correlation between the stall onset timing and the load fluctuation for the low levels of unsteadiness in the slow continuous ramp-up manoeuvre, suggesting that the motion is truly quasi-steady and does not influence the onset of flow detachment.
The mean reaction time of \num{32} convective times observed for this manoeuvre can be considered to be a lower bound for the waiting time between subsequent angle of attack steps in static stall measurements.
For the step manoeuvre, we observe a wider range of fluctuation levels in the holding stage, reaching up to \SI{3}{\percent} of local mean lift coefficient (\cref{fig:tmaxflucts}b).
The reaction time linearly decays with load fluctuations, supporting the fact that flow unsteadiness promotes the onset of stall.
The error bars represent the difference between results obtained with the narrower (\SIrange{6}{94}{\percent}) and the wider threshold limits (\SIrange{2}{98}{\percent}) identifying the start and end of the drop stages.
The error bars are small and indicate that these observations are not sensitive to the drop stage threshold selection for either kinematic.

\begin{figure}
\centering
\includegraphics{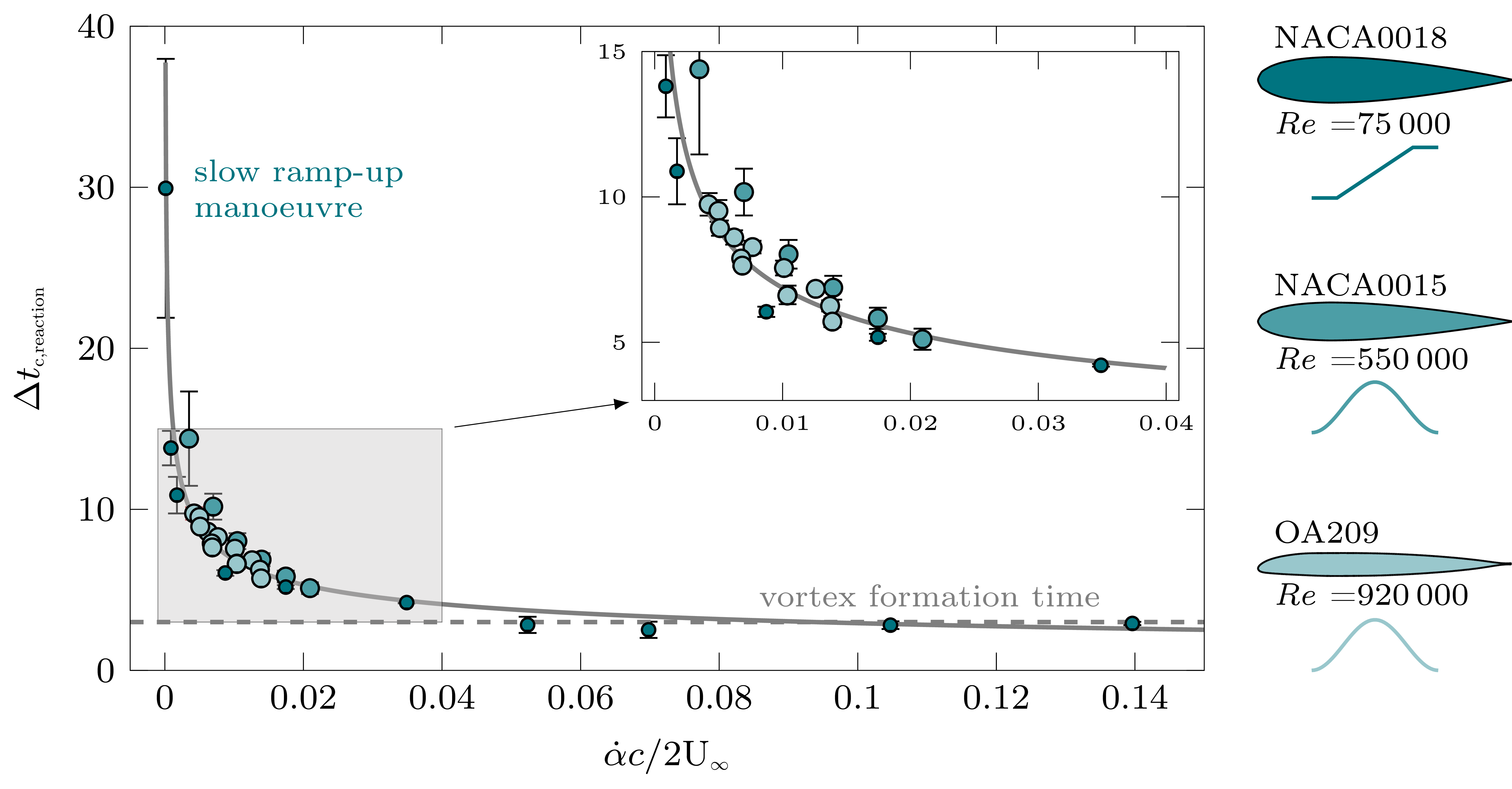}
\caption{Average reaction time delay \kindex{\Delta t}{c,reaction} as a function of reduced pitch rate for three different airfoils.
Results from the continuous ramp-up motion of our NACA0018 are compared with previous results from a sinusoidally pitching OA209 airfoil at \Rey=\num{9.2e5} \cite{Mulleners2012} and a sinusoidally pitching NACA0015 airfoil at \Rey=\num{9.2e5} \cite{He2020}.
}
\label{fig:tmax_pr}
\end{figure}

The relaxation time $\kindex{\Delta t}{c,relax}$ is compared to the load fluctuations in the drop stage in \cref{fig:tmax_pr}.
Both experiments show similar fluctuation levels and a clear decreasing linear trend suggesting load fluctuations occurring during the lift drop promote full flow detachment.
The decrease happens at the same rate for both manoeuvres.
This suggests that the onset of vortex shedding is independent of kinematics for static motions.
This timescale shows a much greater sensitivity to the drop stage threshold selection for some cases.
The increased sensitivity is due to the greater fluctuations that occur in the post-stall regime compared to the pre-stall regime.
The decay in relaxation time with increasing load fluctuations in the drop stage is nevertheless apparent.

The influence of pitch rate on the reaction time was investigated by comparing results of the slow continuous ramp-up manoeuvre with higher constant pitch rate manoeuvres.
Data was collected for reduced pitch rates ranging from \numrange{1e-4}{0.14} with \num{5} repetitions for each pitch rate.
We systematically compute the reaction time $\kindex{\Delta t}{c,reaction} = \kindex{t}{c,pre} -  \kindex{t}{c,ss}$, where $\kindex{t}{c,pre}$ is the time where the lift coefficient starts to drop and $\kindex{t}{c,ss}$ is when the blade exceeds its critical stall angle of \ang{14.2}.
The average reaction time calculated over the \num{5} repetitions is presented against reduced pitch rate in \cref{fig:tmax_pr}.
The error bars show the standard deviation and indicate the spread or the width of the distribution of the measured reaction times at a given reduced pitch rate.

The stall delay or reaction time decreases with increasing pitch rate following a power law decrease.
The additional unsteadiness added to the flow at higher pitch rates promotes flow detachment and the onset of stall.
The reaction time decreases rapidly for reduced pitch rates below \num{0.01} and reaches a plateau at \num{3} convective times for reduced pitch rates above \num{0.04}.
This lower limit is of the order of the vortex formation time of the dynamic stall vortex, which is a classical hallmark of the transition from an attached to a massively separated flow \cite{Mulleners2013,Eldredge:2018cu}.
The plateau represents a minimum timescale the blade requires to form a leading edge vortex and reach a fully separated flow condition \cite{Dabiri2009}.
Stall onset and vortex formation are characterised by the reaction time and it is the main difference used to distinguish static from dynamic stall \cite{Mulleners2012, Eldredge:2018cu}.
The coherent transition between the lowest and higher pitch rates support our hypothesis that the static and dynamic stall responses are phenomenologically the same and their timescales vary continuously as a function of the pitch rate of the underlying motion kinematics.

The standard deviation also decreases rapidly with increasing reduced pitch rate.
The airfoil kinematics play a lesser role in the flow development at extremely low pitch rates and do not longer promote the occurrence of stall, resulting in an increasingly random and wide distribution of the reaction time delays.
The reaction time histogram for the lowest pitch rate (\cref{fig:tmaxhists}) followed a perfect normal distribution, suggesting this motion can be considered as truly quasi-static.

The universality of these results is challenged by comparing them with measurement from different airfoil geometries, kinematics, and Reynolds number.
The timescales of the NACA0018 are compared with those obtained for an OA209 airfoil \cite{Mulleners2012} and for a NACA0015 \cite{He2020} undergoing sinusoidal pitching motions.
As the pitch rate varies continuously for a sinusoidal motion, we use the instantaneous pitch rate at the time the static stall angle is exceeded as the representative effective pitch rate for the sinusoidal motions \cite{Mulleners2012,Deparday2019, Kissing2020}.
The effective pitch rates for the sinusoidal motions vary between \num{0.0035} and \num{0.02}.
The measured stall delays or reaction times for the three different airfoils, subjected to different kinematics, at different Reynolds numbers all collapse onto the same power law decay.
This suggest that stall onset timescales are universal for airfoils undergoing stall at moderate to high \Rey where trailing edge stall is most common.
Further investigations are desirable to explore the ranges of validity of this seemingly universal behaviour in terms of Reynolds number and variety of airfoil geometry.

\begin{figure}
	\centering
	\includegraphics{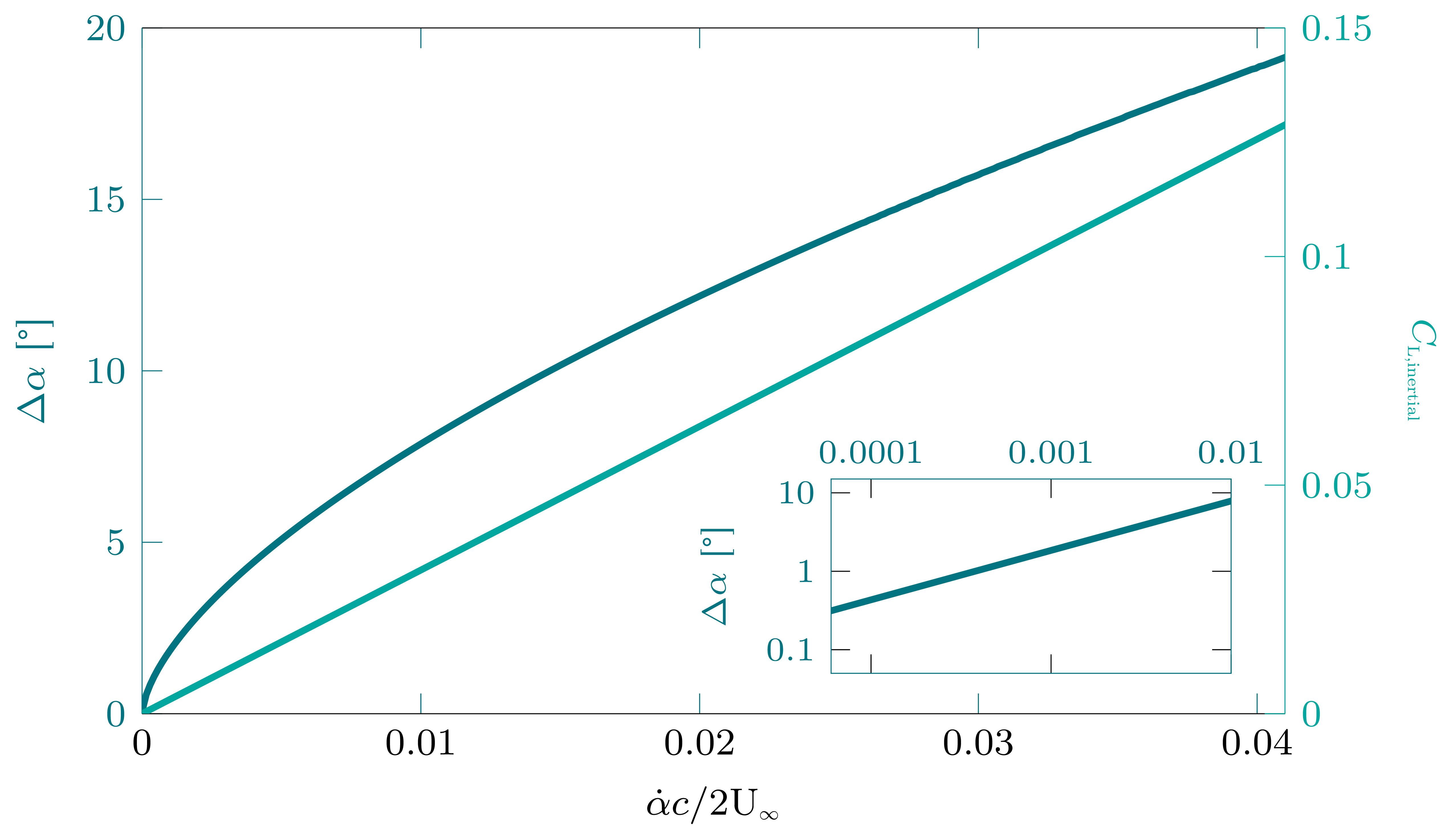}
	\caption{Estimated static stall accuracy $\Delta \alpha$ and inertial component of the lift coefficient $\kindex{C}{L,inertial}$ against reduced frequency for static stall measurements performed with a continuous and uniform ramp-up motion.}
	\label{fig:ssguide}
\end{figure}

The generality of the variation of the stall onset timescales as a function of the unsteadiness of the pitching motion presented in \cref{fig:tmax_pr} can be used to lay out guidelines for reliably measuring the static stall angle and lift and drag polars.
The systematic acceleration and deceleration related to a stepwise increase of the angle of attack is more likely to disturb the flow than a continuous motion.
A continuous ramp-up motion with slow uniform pitch rate is thus preferred but how slow is slow enough?
To answer that, we first fit a generalised power law decay to our experimental data yielding the following expression:
\begin{equation}\label{eq:powerdecay}
\kindex{\Delta t}{c,reaction} = 1.25\left(\frac{\dot{\alpha}c}{2\Uinf}\right)^{-0.37} \quad.
\end{equation}
This expression is used to determine the angular accuracy for the measurement of the static stall angle for a given pitch rate determined by the angular increase that occurs during expected stall reaction time:
\begin{equation}
\Delta \alpha = \frac{\dot{\alpha}c}{2U_\infty} 2\kindex{\Delta t}{c,reaction} = 2.50\left(\frac{\dot{\alpha}c}{2\Uinf}\right)^{0.63}\quad.
\end{equation}
The motion should be slow enough to minimise $\Delta \alpha$.
In addition, we want to limit the inertial lift contributions associated with a dynamic motion.
The inertial contribution to the lift coefficient for a continuous ramp-up motion can be estimated based on Theodorsen's theory \cite{theodorsen} as:
\begin{equation}
\kindex{C}{l,inertial} = \frac{\pi\dot{\alpha}c}{2\Uinf}\quad.
\end{equation}
The evolution of both the static stall angular accuracy $\Delta \alpha$ and the inertial component of the lift coefficient $\kindex{C}{l,inertial}$ as a function of the reduced pitch rate are presented in \cref{fig:ssguide}.
Overall, the quasi-steady inertial lift contributions are a lesser issue than the stall angle increases.
Reduced frequencies of the order of \num{1e-4} yield a static stall angle accuracy of $\Delta \alpha<\ang{1}$.
The quasi-steady inertial contribution for these pitch rates are negligible.
When measuring a static lift response using a continuous slow ramp-up motion, the lift response can be considered a conventional static force response, free of unsteady and quasi-steady influences, and providing a reliable estimate of the critical static stall angle for $(\dot{\alpha}c)/(2U_\infty)<\num{1e-4}$.

\section{Conclusion}
We investigated the dynamic load variations and timescales of static stall by measuring loads acting on a blade undergoing two quasi-steady manoeuvres:
\begin{inparaenum}[(i)]
\item a slow continuous ramp-up motion from angles of attack \ang{8} to \ang{18} at a constant reduced pitch rate of \num{1.3e-4} and
\item a step-wise increase in angle of attack from \ang{14.2} to \ang{14.8} within \num{0.04} convective times.
\end{inparaenum}
We defined three characteristic time delays associated with the transient flow development from attached to fully separated in response to the two types of manoeuvres: a reaction delay, a drop delay, and a relaxation delay.

The reaction delay is the time delay between the moment when the blade exceeds its stall angle of \ang{14.2} and the moment when the lift collapses.
This timescale characterises the duration of stall onset, which plays a central role in the distinction between static and dynamic stall.
The reaction time is not influenced by pre-stall fluctuations for the slowest continuous ramp-up motion, and its occurrence histogram follows a normal distribution centred around \num{32} convective times.
The unsteadiness induced by the step manoeuvre leads to pre-stall load fluctuations that are three times larger than those induced by the slow continuous ramp-up manoeuvre and do promote the onset of stall.
The reaction delay linearly decreases with increasing fluctuations in the step manoeuvre.

The results for the reaction delay from the slow continuous ramp-up motion were compared with results from dynamic ramp-up manoeuvres with reduced pitch rates ranging from \numrange{1.3e-4}{0.14} and with previously obtained results from dynamic sinusoidal pitching motions with different airfoil geometries at different Reynolds numbers.
This comparison revealed a universal power law decay of the stall delays from \num{32} convective times for the lowest pitch rates to a plateau around \num{3} convective times for reduced pitch rates above \num{0.04}.
The plateau level matches the vortex formation time, which is the minimum time interval required for the boundary layer to roll-up into a coherent stall vortex and separate from the airfoil.
The standard deviation of the observed stall delays across multiple repetitions also rapidly decreased with increasing pitch rate which aids in promoting the occurrence of stall.
Static stall is not phenomenologically different than dynamic stall and is merely a typical case of stall for low pitch rates where the onset of flow separation is not promoted by the blade kinematics.

Based on the results, we propose that conventional static stall polars should be measured using a continuous and uniform ramp-up at a reduced frequency $<\num{1e-4}$ to minimise the angle of attack variation during the stall delay.
The inertial lift contributions at these pitch rates are negligible.
A continuous motion is preferred to a stepwise increase as the systematic acceleration and deceleration of a step motion is more likely to cause an unsteady flow response.
If a step-wise motion is selected, it is advised to wait at least \num{30} convective times between the end of the step and the start of the measurements to allow for the flow to respond to the change in the angle of attack.


\section*{Acknowledgements}																		%
This work was supported by the Swiss national science foundation under grant number PYAPP2\_173652.

\bibliography{doss_v2}%

\end{document}